\documentclass[twocolumn,showpacs,preprintnumbers,amsmath,amssymb]{revtex4}

\usepackage{graphicx}
\usepackage{dcolumn}
\usepackage{bm}

\begin{document}


\title{Current-controlled dynamic magnonic crystal}

\author{A. V. Chumak}

\email{chumak@physik.uni-kl.de}

\author{T. Neumann}

\author{A. A. Serga}

\author{B. Hillebrands}

\affiliation{Fachbereich Physik  and Forschungszentrum OPTIMAS,
Technische Universit\"at Kaiserslautern, 67663 Kaiserslautern,
Germany}

\author{M. P. Kostylev}
\affiliation{School of Physics, University of Western Australia, Crawley, Western Australia 6009,
Australia}

\date{\today}

\begin{abstract}
We demonstrate a current-controlled, dynamic magnonic crystal. It consists of a ferrite film whose
internal magnetic field exhibits a periodic, cosine-like variation. The field modulation is created by a
direct current flowing through an array of parallel wires placed on top of a spin-wave waveguide. A
single, pronounced rejection band in the spin-wave transmission characteristics is formed due to
spin-wave scattering from the inhomogeneous magnetic field. With increasing current the rejection band
depth and its width increase strongly. The magnonic crystal allows a fast control of its operational
characteristics via the applied direct current. Simulations confirm the experimental results.
\end{abstract}


\maketitle

Spin waves in magnetic materials attract special attention because of their potential application as
information units in signal processing devices. Digital spin wave logic devices \cite{logic, Lee08} as
well as devices for analogous signal processing \cite{Ada88, convolver, PRL-restoration} can be
fabricated based on spin waves. It has been shown that the spin-wave relaxation, one of the main
obstacles for spin-wave application, can be overcome by means of parametric amplification \cite{Sch60,
PRL-restoration}.

The study of spin waves in magnetic materials is also interesting from a fundamental point of view. The
interaction of the numerous existing spin-wave modes in ferromagnetic samples \cite{Kalinikos86} as well
as nonlinear effects such as soliton formation \cite{Kal83, m-soliton} are just some examples.

Magnonic crystals, which are defined as artificial media with a spatially periodic variation of some of
their magnetic parameters, constitute a research field which connects fundamental physics with
application \cite{ MC-review, Gul03, Gubbiotti1, APL-Chumak, Wan09}. They are the analogue of photonic
crystals which operate with light. The spectra of spin-wave excitations in such structures are
considerably modified compared to uniform media and exhibit features such as full band gaps where spin
waves are not allowed to propagate.

Promising functionalities arise by taking advantage of the dynamic controllability and by potentially
changing the characteristics of the magnonic crystal faster than the spin-wave relaxation time: even the
simple possibility to "switch" a periodical inhomogeneity on and off immediately offers a method to trap
and release a spin wave packet. This can be exploited for instance in information storage.

Here, we present a first realization of such a dynamic magnonic crystal. It is based on spin-wave
propagation in an yttrium iron garnet (YIG) film placed in a periodically varying, dynamically
controllable magnetic field. The magnetic field is created by the superposition of a spatially
homogeneous bias magnetic field with the localized Oersted fields of current carrying wires which are
placed in an array layout close to the YIG film surface \cite{Fet89}. By controlling the direct current
in the wires the field modulation is adjusted and the spin-wave transmission is changed from full
transmission for no applied current to a transmission showing a distinct, $30~{\rm MHz}$-wide stop band
for an applied current of $1.25~{\rm A}$. The dynamic controllability constitutes a major difference to
previous realizations of magnonic crystals with a periodically varying magnetic field \cite{Vor88}.

Previous studies focused on the interaction of propagating spin-wave packets with the Oersted field of a
single current carrying wire or a set of two wires at most \cite{Dem04, Kos07, Han07, Neu09, Ser09}. It
was shown that the spin-wave transmission can be effectively changed by varying the value of the direct
current. However, the appearance of a pronounced frequency stop-band, for which spin-wave transmission
is prohibited (while it remains almost unaffected outside the band), is only observed for larger wire
numbers.

\begin{figure}
\includegraphics[width=0.8\columnwidth]{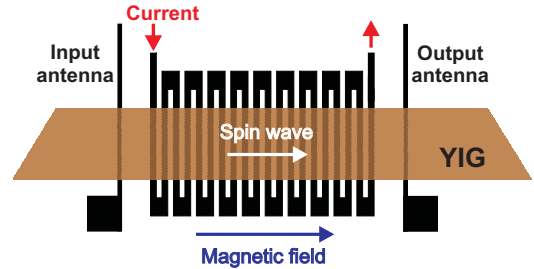}
\caption{\label{setup} (Color online) Sketch of the magnonic crystal structure used in the experiments.}
\end{figure}

A sketch of the experimental section is shown in Fig.~\ref{setup}. It consisted of a $5~\mu{\rm
m}$-thick YIG film which was epitaxially grown on a gallium gadolinium substrate. A bias magnetic field
of $4~\pi\cdot1.6~{\rm A m}^{-1}$ was applied along the YIG waveguide so that the conditions for the
propagation of backward volume magnetostatic waves (BVMSWs) are given.

To achieve a periodic modulation of the magnetic field an array of connected, parallel wires was
designed. The wire structure was patterned by means of photolithography on an aluminium nitride
substrate with high thermal conductivity in order to avoid heating. The structure consists of 40 wires
of $75~\mu{\rm m}$ width with a $75~\mu{\rm m}$ spacing.

The wire array was placed above the YIG film in such a way that the wires ran perpendicularly to the
spin-wave waveguide. Thus, the magnetic Oersted field produced by each of the current carrying wire
segments is oriented in first approximation parallel to the bias magnetic field.

In the experiment the individual wires were connected to form a meander structure \cite{Fet89} where the
current in neighboring wires flows in opposite directions (see Fig.~\ref{setup}). Thus, a magnonic
crystal with a lattice constant $a = 300~\mu{\rm m}$ and 20 repetitions was fabricated.

Another possible configuration would have all currents flowing in the same directions ("multi-strip
structure") so that for all wires the Oersted fields have identical orientation. This is an advantage
since our previous studies have shown the existence of two physically different regimes for the
different field orientations: When the Oersted field decreases the internal field one implements the
spin-wave tunneling regime \cite{Dem04}. When the internal field is locally increased the conditions for
resonant spin-wave scattering \cite{Kos07} can be fulfilled for which the spin-wave transmission depends
non-monotonically on the applied current and exhibits a strong frequency dependence \cite{Neu09}.

However, the meander structure has important advantages: (i) It produces a much stronger field
modulation because the in-plane components of the Oersted fields for neighbouring wires are oriented in
opposite directions. (ii) It ensures that the magnetic field averaged over the structure remains
constant for any current magnitude. 

Two microstrip antennas were placed, one in front and another one behind the wire structure (see
Fig.~\ref{setup}) in order to excite and detect BVMSWs. A network analyzer connected to the input and
output antennas was used to measure the spin-wave transmission characteristics.

In order to minimize the electromagnetic coupling between the current carrying wire segments and the
spin waves, a $100~\mu{\rm m}$ thick SiO$_2$ spacer was placed between the YIG film and the wire
structure.
Note, that the spin-wave dipole field decays exponentially with the distance form the film
surface while the Oersted field around the wires scales with the inverse distance between the wire and
the film surface. The distance of 100 microns between the wire array and the film surface proved to be
large enough to avoid any disturbance of the spin-wave propagation by the meander conductor, but still
small enough to ensure an efficient modulation of the magnetic field in the film by the current field.

\begin{figure}
\includegraphics[width=0.95\columnwidth]{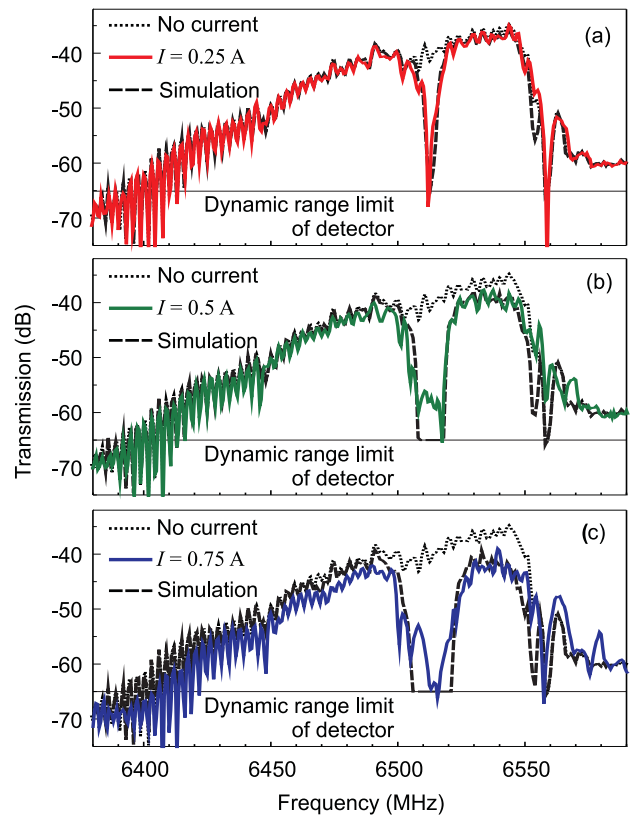}
\caption{\label{transmission} (Color online) Spin-wave transmission characteristics in a uniform
magnetic field (dotted curves) and in a magnetic field which is periodically modulated by the current
$I$ applied to the wires (solid curves). The dashed curves are calculated as the product of the
experimentally observed transmission in a uniform field with the transmission coefficient obtained from
the numerical simulations (where in addition the limited dynamic range of the experimental setup is
taken into account).}
\end{figure}

Experimental results are shown in Fig.~\ref{transmission}. The dotted curves in the panels show the
transmission characteristics without direct current applied to the wires. They are typical for BVMSWs,
limited by the ferromagnetic resonance frequency towards high frequencies and by the antenna excitation
efficiency from the opposite side. The minimal transmission loss of about 35~dB is determined by the
spin-wave excitation/reception efficiency of the microwave antennas and by the spin-wave relaxation
parameter of the ferrite film.

Figure~\ref{transmission}(a) shows that the application of a current $I=0.25~{\rm A}$ to the structure
results in the appearance of a pronounced rejection band at a frequency $f_1 \approx 6510~{\rm MHz}$
where the transmission of spin waves is prohibited. The rejection band already appears for a current as
small as $80~{\rm mA}$. With an increase in the current the rejection band depth increases rapidly and
for $0.25~{\rm A}$ it reaches the dynamic range of the experimental setup which is limited mainly  by
the direct electromagnetic leakage between the microstrip antennas. A further increase in the current
applied to the wires results in a pronounced broadening of the rejection band (see
Fig.~\ref{transmission}(b) and Fig.~\ref{transmission}(c)).

We emphasize one particularly interesting feature of the magnonic crystal presented here. Practically
only one rejection band is formed. This is not true for magnonic crystals consisting of an array of
grooves on the YIG film surface \cite{APL-Chumak} where multiple rejection bands are formed. As shown by
our calculations, for the magnonic crystal studied here the spatially periodic modulation of the
magnetic field is close to cosinusoidal. For an ideal harmonic variation only one rejection band should
exist since the reflection amplitude is proportional to the Fourier component of the inhomogeneity
profile corresponding to twice the spin-wave wave vector as seen from Eq. (6) in \cite{Kos07}.

\begin{figure}
\includegraphics[width=0.98\columnwidth]{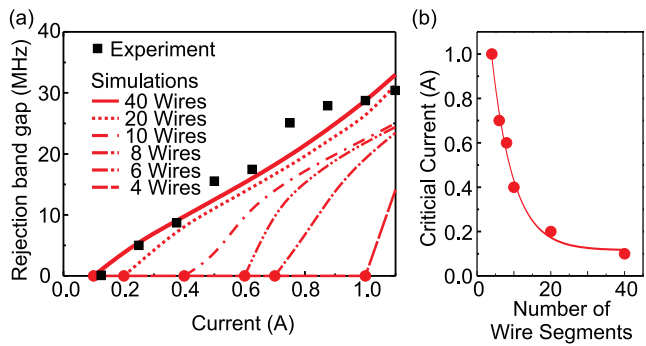}
\caption{\label{width} (Color online) (a) Experimentally obtained width of first rejection band as a
function of the current applied to the wires. The lines and filled circles represent the results of
numerical simulations. (b) Calculated critical current for which a band gap with a suppression $\geq
90\%$ is obtained for different numbers of parallel wire segments. The line indicates a fit with a
decaying exponential function.}
\end{figure}


The presence of only one rejection band is an advantage for applications in a microwave filter device.
Another positive aspect of the presented dynamic crystal is that practically no losses occur for
frequencies outside the induced stop band with increasing current (see Fig.~\ref{transmission}(c)). In
the groove-structure-based magnonic crystal \cite{APL-Chumak} such undesired parasitic increase of
losses in the transmission bands was observed for larger groove depths.

In order to investigate the dynamic properties of the magnonic crystal additional experiments with a
pulsed direct current supplied to the meander structure were performed. The microwave frequency which
was applied to the input antenna to excite the spin-wave signal was chosen inside the rejection band ($f
= 6.51~{\rm GHz}$). The direct current supplied to the meander structure was pulsed with a duration of
$50~{\rm ns}$ and a strength of $0.5~{\rm A}$. The obtained results show that the spin-wave transmission
can be dynamically turned on and off with a transition time for the magnonic crystal of approximately
$50~{\rm ns}$.

The measured rejection band width as a function of the applied current is shown in Fig.~\ref{width}(a).
It was measured for the first rejection band at the power level where the spin-wave intensity decreases
to one tenth, i.e. $-10~{\rm dB}$, of its value. One can see that the band width can be tuned from
$5~{\rm MHz}$ for $125~{\rm mA}$ current to $31~{\rm MHz}$ for $1.25~{\rm A}$ and exhibits a linear
behavior with respect to the applied current. The possibility of a dynamical control of the rejection
band width seems to be promising for the design of a dynamic stop-band microwave filter. The center
frequency of the rejection band can be controlled by means of an applied magnetic field.

The experimental results were confirmed by numerical simulations. We used a 1-dimensional approach, in
which the dipole field was expressed via a Green's function and the magnetic field was averaged over the
film thickness. Details on the model can be found in \cite{Kos07}. The obtained frequency-dependent
transmission curves show a single, well pronounced stop-band for low currents which coincides well with
the experiment (see Fig.~\ref{transmission}). The calculated rejection efficiency, given by the depth of
the stop band, reached up to $-150~{\rm dB}$ which exceeds the dynamic range in the experiment greatly.

Two effects are observed if the number of parallel wire segments in the simulation is decreased:
Firstly, the achieved stop-band width for a given current decreases slightly. Secondly, the rejection
efficiency decreases dramatically. As a consequence, the desired signal suppression (e.g. one tenth of
the transmission for no applied current) is only reached for higher currents which results in the
behavior of the rejection band gap width shown in Fig.~\ref{width}(a). Figure~\ref{width}(b) summarizes
the calculated dependence of the critical current necessary to obtain a $10~{\rm dB}$ signal suppression
on the number of parallel wire segments. As can be seen a larger number of wires exponentially reduces
the critical current which has a particularly pronounced effect on the current if the number of wires is
smaller than 20.

In conclusion, we presented a current-controlled magnonic crystal whose operational characteristics can
be tuned dynamically within a transition time of $50~{\rm ns}$. The spatially periodic Oersted field of
a meander conductor located in the vicinity of YIG film surface results in a pronounced modification of
spin wave dispersion which leads to the appearance of spin-wave rejection bands. The width of the main
rejection band varies linearly with the magnitude of applied direct current and can be tuned in the
range from $5~{\rm MHz}$ to $30~{\rm MHz}$. Numerical simulations are in good qualitative agreement with
the experiment. Overall, the presented dynamic magnonic crystal is promising for the investigation of
linear and nonlinear spin-wave dynamics and can be used as a dynamically controlled microwave stop-band
filter.

Financial support by the DFG project SE 1771/1-1, the Matcor Graduate School of Excellence, the
Australian Research Council, and the University of Western Australia is acknowledged. Special
acknowledgments go to the Nano+Bio Center, TU Kaisers\-lautern. T. Neumann would like to thank
especially Robert L. Stamps and the University of Western Australia for their assistance during his
research stay.

\end{document}